\documentclass[10pt]{article}   	
\usepackage{geometry}                		
\usepackage{amssymb}
\usepackage{bm}
\usepackage{cite}

\usepackage{cite}
\usepackage[pdftex]{graphicx}
\usepackage{subfigure}

\title{Annealing Approach to Quantum Tomography\footnote{This work is based on results from a project commissioned by the new Energy and Industrial Technology Development Organization (NEDO), Japan}}
\author{Kentaro Imafuku\\
National Institute of Advanced Industrial Science and Technology (AIST)\\
Aomi 2-3-26, Koto-ku, Tokyo 1350064, Japan}
\date{\empty}							

\begin{document}
\maketitle
\begin{abstract}
Annealing approach to quantum tomography is theoretically proposed. First, based on the maximum entropy principle, we introduce classical parameters to combine ``quantum models (or quantum states)" given a prior for potentially representing the unknown target state. Then, we formulate the quantum tomography as an optimization problem on the classical parameters, by employing relative entropy of the parametrized state with the target state as the objective function to be minimized. We show that the objective function is physically implementable, in a theoretical sense at least, as an effective Hamiltonian to be induced by physical interactions of the system with environment systems being prepared in the target state. Corollary, applying quantum annealing to the effective Hamiltonian, we can execute quantum tomography by obtaining the ground state that gives the optimal parameters.
\end{abstract}
\section{Introduction}
Quantum tomography\cite{PhysRevA.45.7688,PhysRevA.53.2078,PhysRevA.40.2847,PhysRevLett.70.1244,Hradil2004,Dariano2004} is a fundamental process obtaining information from a set of an identical but unknown quantum states to approximately estimate the state. (We call the state to be estimated ``target state" through this article.) We consider an application of annealing computation\cite{PhysRevE.58.5355,2000quant.ph..1106F,Farhi2000ANS,RevModPhys.80.1061,nature10012,2014McGeoch} to quantum tomography with which we directly obtain the estimation as outputs of the computation. In our approach, the set of the target states is used as environment systems inducing an effective Hamiltonian with a ground state giving ``representation" of the target state. Applying the quantum annealing computation to the effective Hamiltonian, we can execute a quantum tomography without humanly processing including quantum measurements on the target state, except for the final read out of the result of the computation. As an introduction, we give a sketch of our idea below. Let $\hat{\mu}$ on Hilbert space ${\mathcal H}$ be the target state that we like to take the quantum tomography. On the Hilbert space, we define ${\mathcal M}$ as a set of quantum states as
\begin{equation}\label{eq:models}
{\mathcal M}:=\{\hat{\rho}_i\}_{i\in\{1,\cdots,m\}}.
\end{equation}
For a technical reason explained in the following sections, we suppose that each state in ${\mathcal M}$ is of full rank with respect to ${\mathcal H}$, {\it i.e.},
\begin{equation}
\mbox{rank}~\hat{\rho}_i=\dim \mathcal H
\end{equation}
for all $i \in \{1,\cdots,m\}$. Aiming to approximately obtain a representation of $\hat{\mu}$, we consider a quantum state defined with $\hat{\rho}_i \in {\mathcal M}$ with real numbers ${\bm \omega}:=(\omega_0,\omega_1,\cdots,\omega_m)$ as
\begin{equation}\label{eq:parameterization}
\hat{\rho}(\bm{\omega}):=\exp\left[\sum_{i=0}^{m}\omega_i \hat{\eta}_i\right]
\end{equation}
where
\begin{equation}\label{eq:etas}
\hat{\eta}_0:=\hat{I}_{{\mathcal H}},\quad\mbox{and}\quad
\hat{\eta}_i:=-\ln\hat{\rho}_i~\mbox{for all}~i \in \{1,\cdots,m\}
\end{equation}
with $\hat{I}_{{\mathcal H}}$ denoting the identity operator on ${\mathcal H}$. Notice that the above parametrization is based on the maximum entropy principle\cite{PhysRev.106.620,jaynes_justice_1986}. Employing the relative entropy\cite{RevModPhys.74.197} of $\hat{\rho}({\bm \omega})$ with respect to $\hat{\mu}$, {\it i.e.},
\begin{equation}\label{eq:relative entropy}
R\left(\hat{\mu};\hat{\rho}({\bm \omega})\right):={\rm tr}\left(\hat{\mu} \ln\hat{\mu}\right)-{\rm tr}\left(\hat{\mu} \ln\hat{\rho}({\bm\omega})\right)
\end{equation}
or
\begin{equation}\label{eq:objective function D}
D\left(\hat{\mu};\hat{\rho}({\bm \omega})\right):=-{\rm tr}\left(\hat{\mu} \ln\hat{\rho}({\bm\omega})\right),
\end{equation}
as a metric to be minimized, we can map the quantum tomography to the finding problem of the minimizing argument on ${\bm \omega}$. Finally, constructing a Hamiltonian corresponding to the metric with the degree of freedoms of $\hat{\bm \omega}$, we can apply the quantum annealing to the finding argument problem. 

In the following sections, we add some more detailed explanations or formulations on the each part of the above idea, as well as numerical examples.

\section{Maximum Entropy Parameterization}
In this section, we give implications of the parameterization in eq.(\ref{eq:parameterization}). As is well known, the form of eq.(\ref{eq:parameterization}) ensures that $\hat{\rho}({\bm \omega})$ is the maximum entropy state among states with the same expectation values of $\hat{\eta}_i$ for all $i$. For a general state $\hat{\rho}$, the expectation value
\begin{equation}
\eta_i:={\rm tr}\left(\hat{\rho} \hat{\eta}_i\right)=-{\rm tr}\left(\hat{\rho} \ln \hat{\rho}_i\right)
\end{equation}
can be interpreted as the mean code length required in recording an output of the rank $1$ projection measurement designed to make the length optimal under a hypothesis that $\hat{\rho}$ was $\hat{\rho}_i$. Each $\hat{\rho}_i$ in ${\mathcal M}$ in eq.(\ref{eq:models}) is called a model because $\rho_i \in {\mathcal M}$ are introduced under a hypothesis that they can approximately reconstruct the target state $\hat{\mu}$ by being combined in an appropriate manner. One possible way to combine models is the one based on the maximum entropy principle. The idea of the parameterization can be understood as the follows: As the first step, we consider maximum entropy states corresponding to various ${\bm \eta}:=\left(\eta_1,\cdots,\eta_m\right)$ with respect to the models defined by ${\mathcal M}$. Then, among the various maximum entropy states, we pick up one according to a metric. The idea can be regarded as a sort of quantum extension of a concept called ``context mixing" that is often used in reconstructing the unknown target probability distribution by combining some probability distributions given as model a priori\cite{Mahoney2005,Kulekci2011,cmki2019}. Notice that when every $\hat{\rho}_i \in {\mathcal M}$ commutes each other, our parametrization simply returns to the original parameterization used in the classical context mixing problem.

\subsection*{Example}
\begin{figure}[h]
\centering
\subfigure[$\theta=-\pi/2$]{
\includegraphics[width=.7cm]{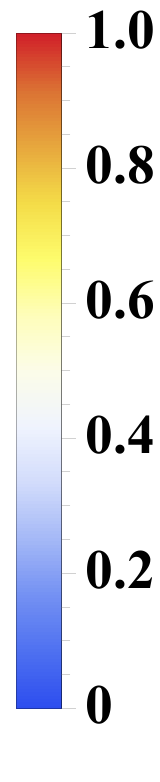}
\includegraphics[width=3cm]{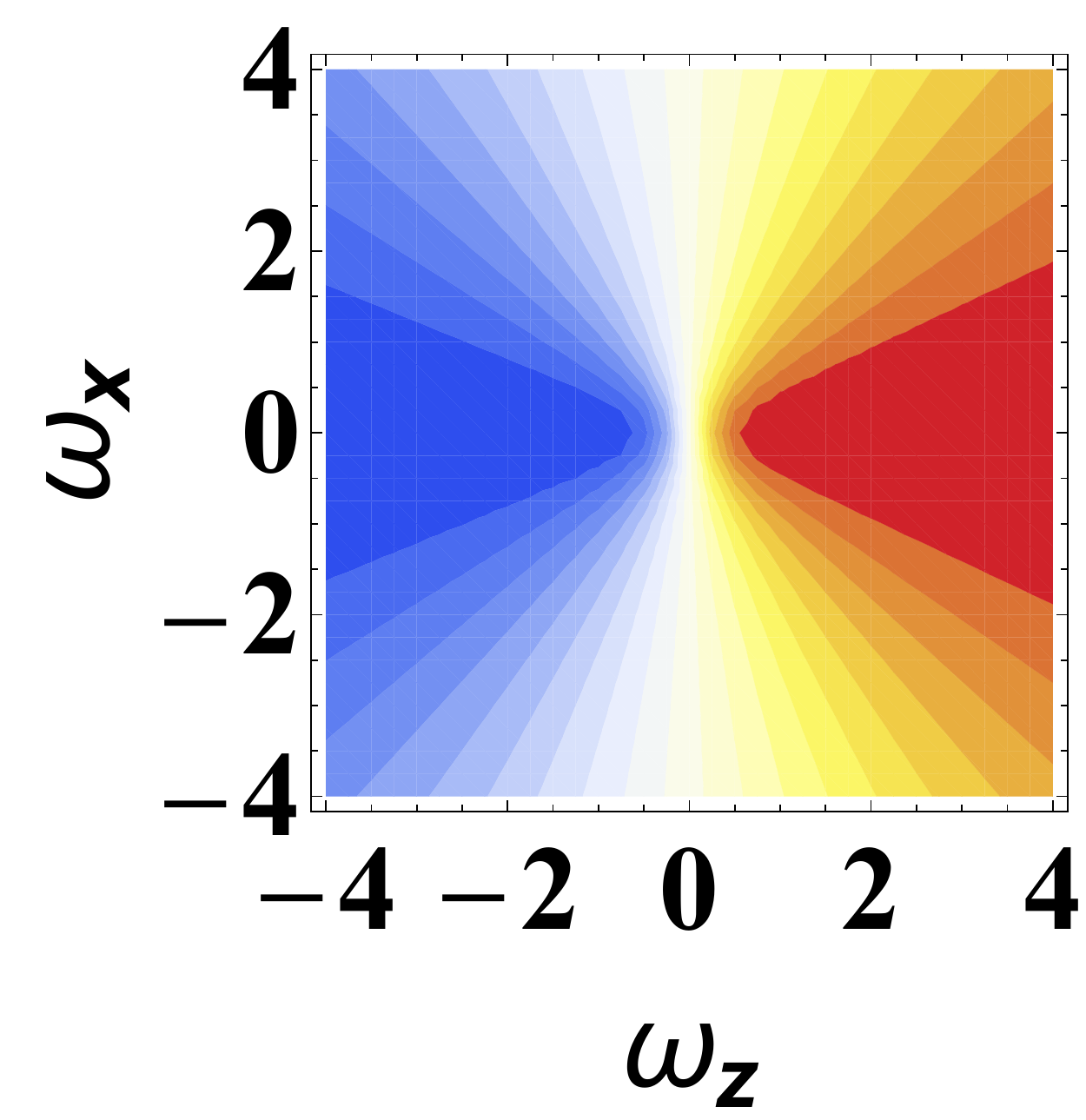}
\label{fig:1a}}
\subfigure[$\theta=-3\pi/8$]{
\includegraphics[width=3cm]{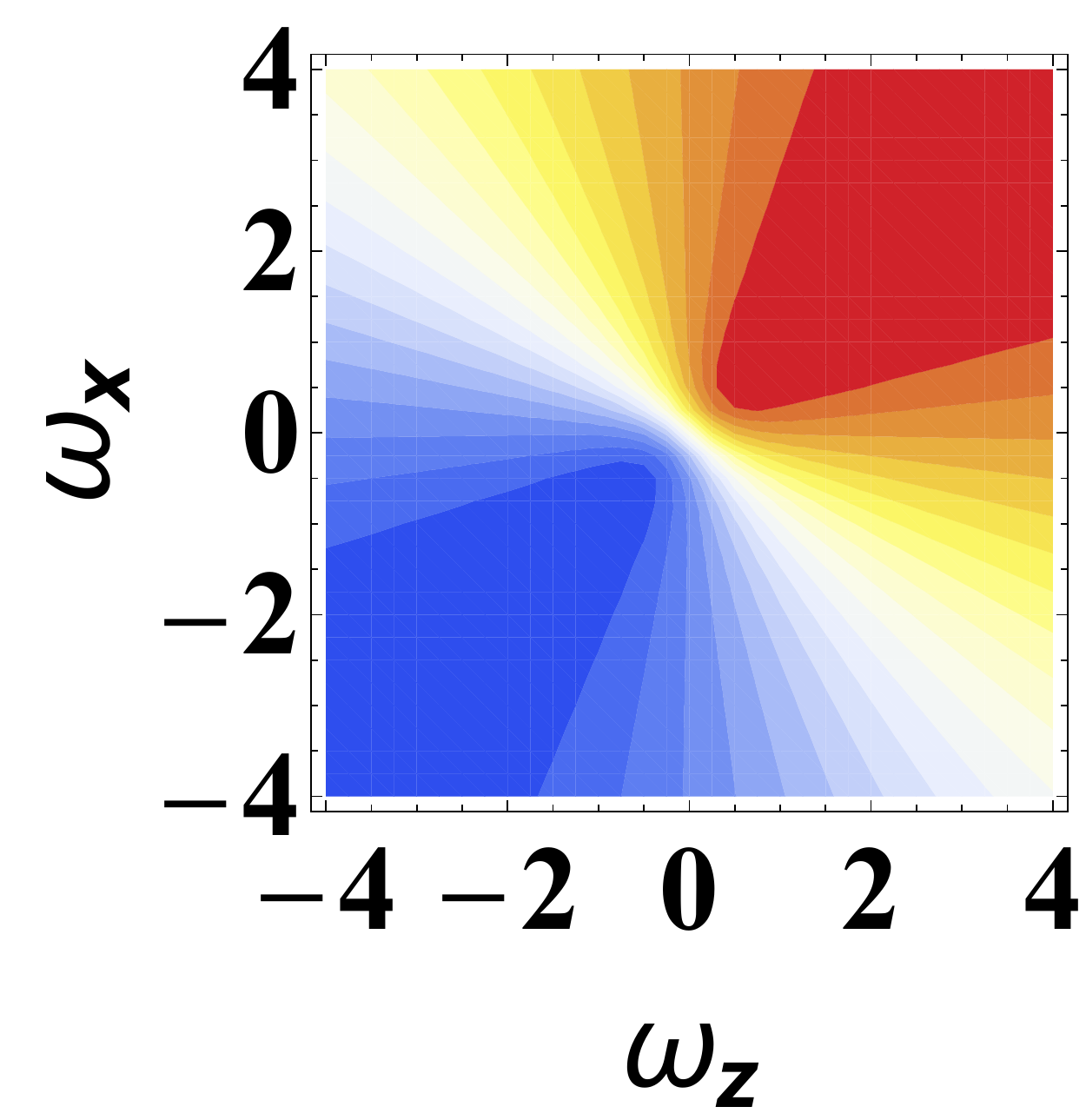}
\label{fig:1b}}
\subfigure[$\theta=-\pi/4$]{
\includegraphics[width=3cm]{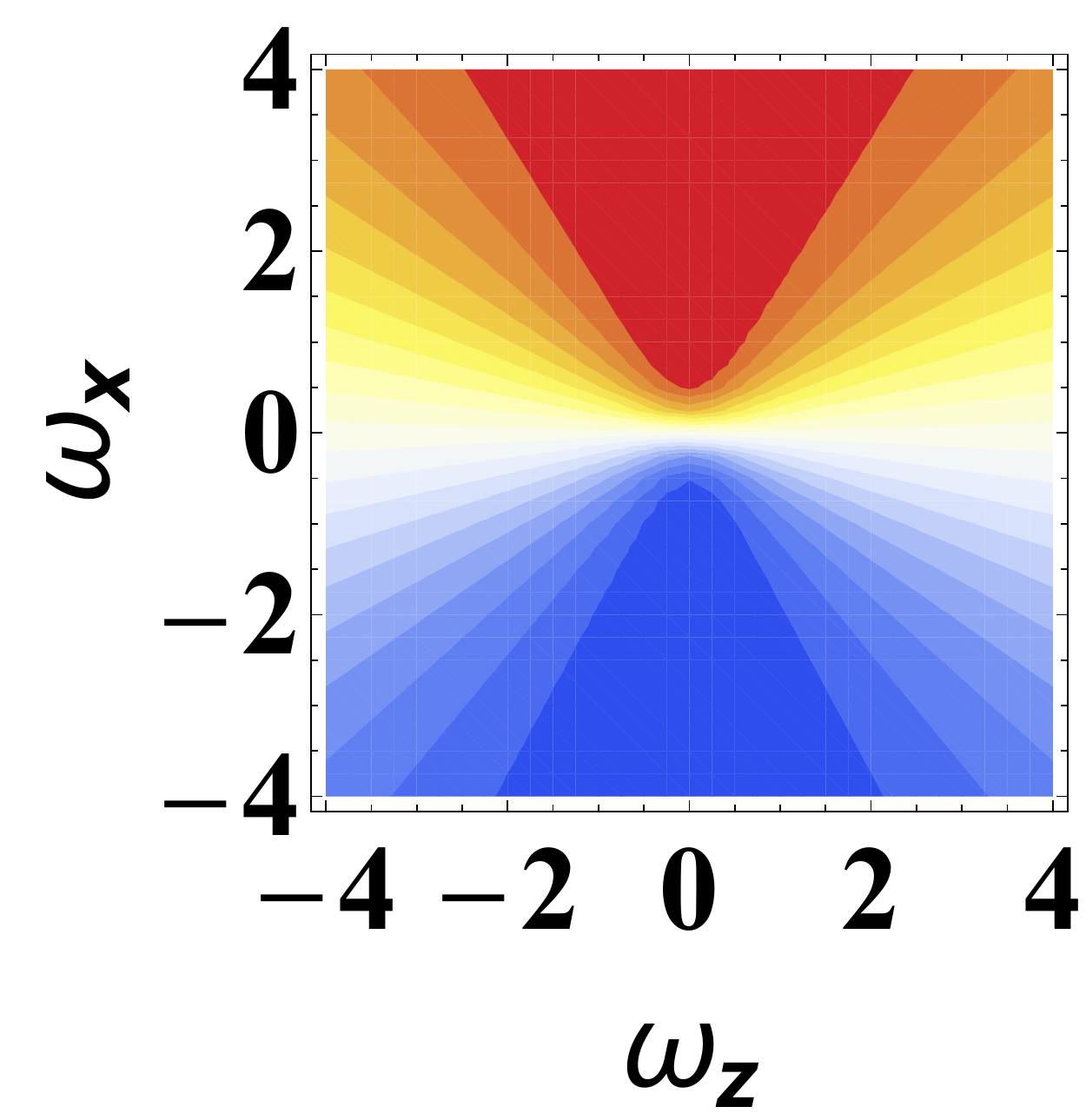}
\label{fig:1c}}
\subfigure[$\theta=-\pi/8$]{
\includegraphics[width=3cm]{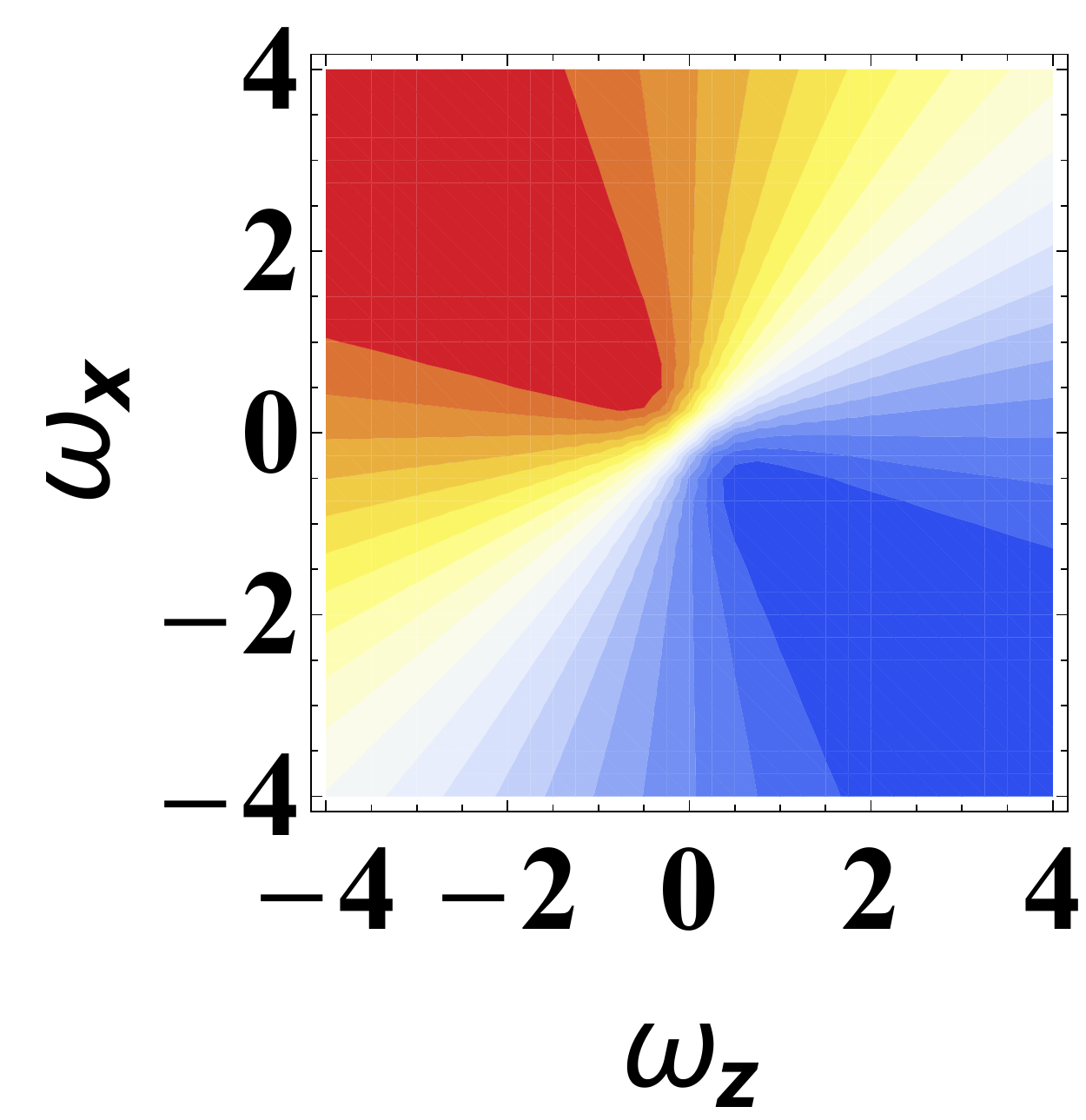}
\label{fig:1d}}\\
\subfigure[$\theta=0$]{
\includegraphics[width=.7cm]{./legend0.pdf}
\includegraphics[width=3cm]{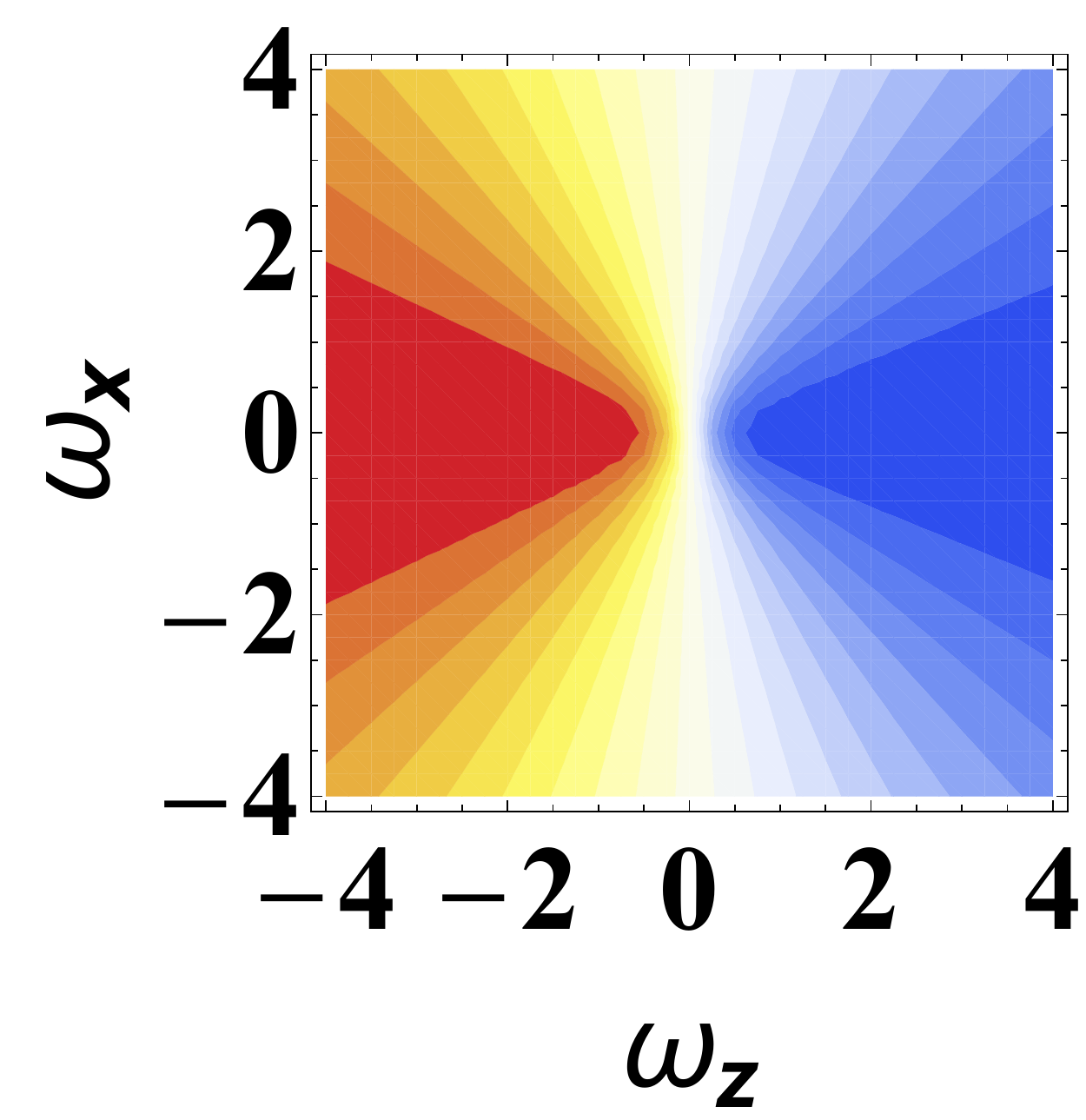}
\label{fig:1e}}
\subfigure[$\theta=+\pi/8$]{
\includegraphics[width=3cm]{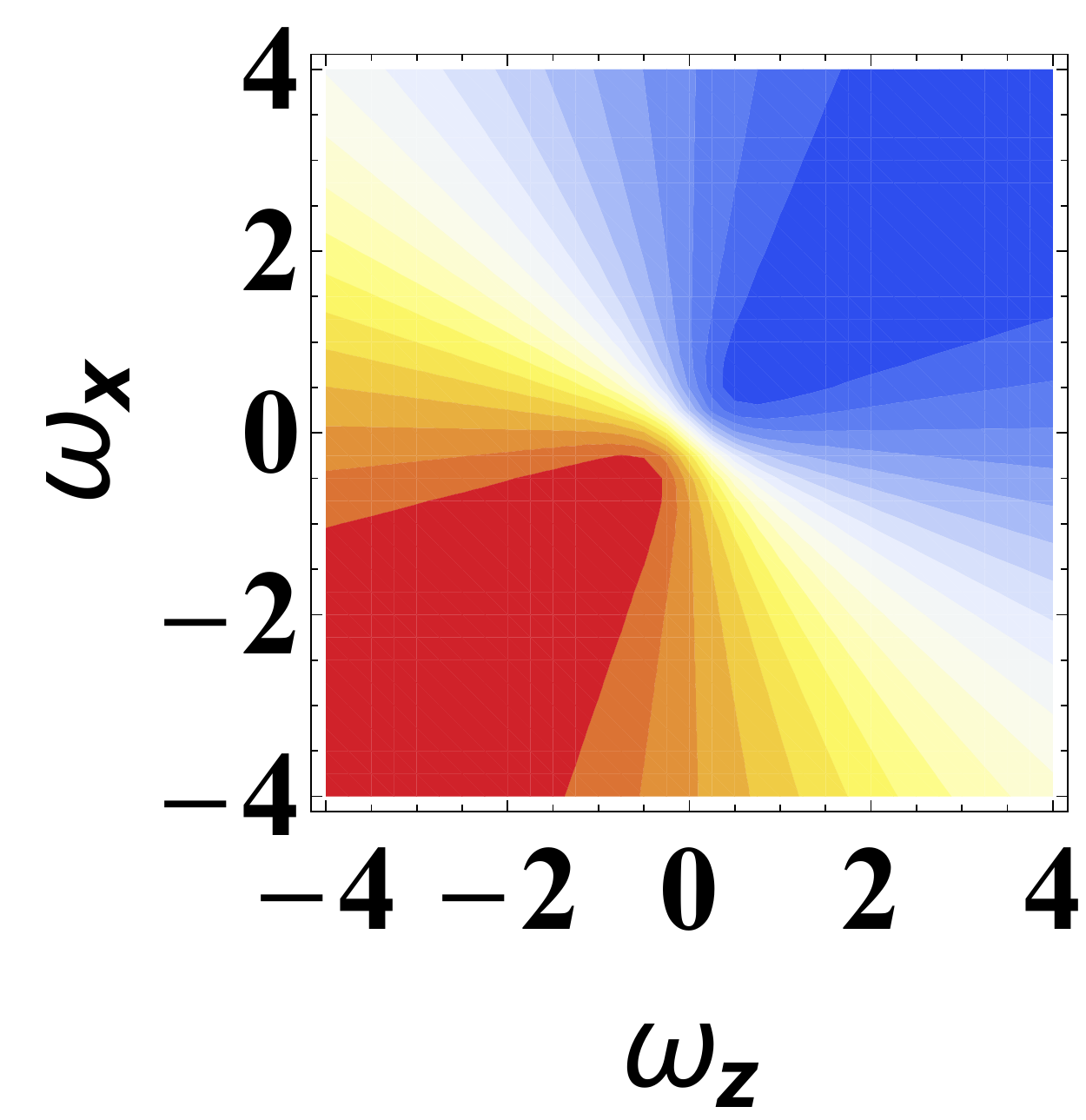}
\label{fig:1f}}
\subfigure[$\theta=+\pi/4$]{
\includegraphics[width=3cm]{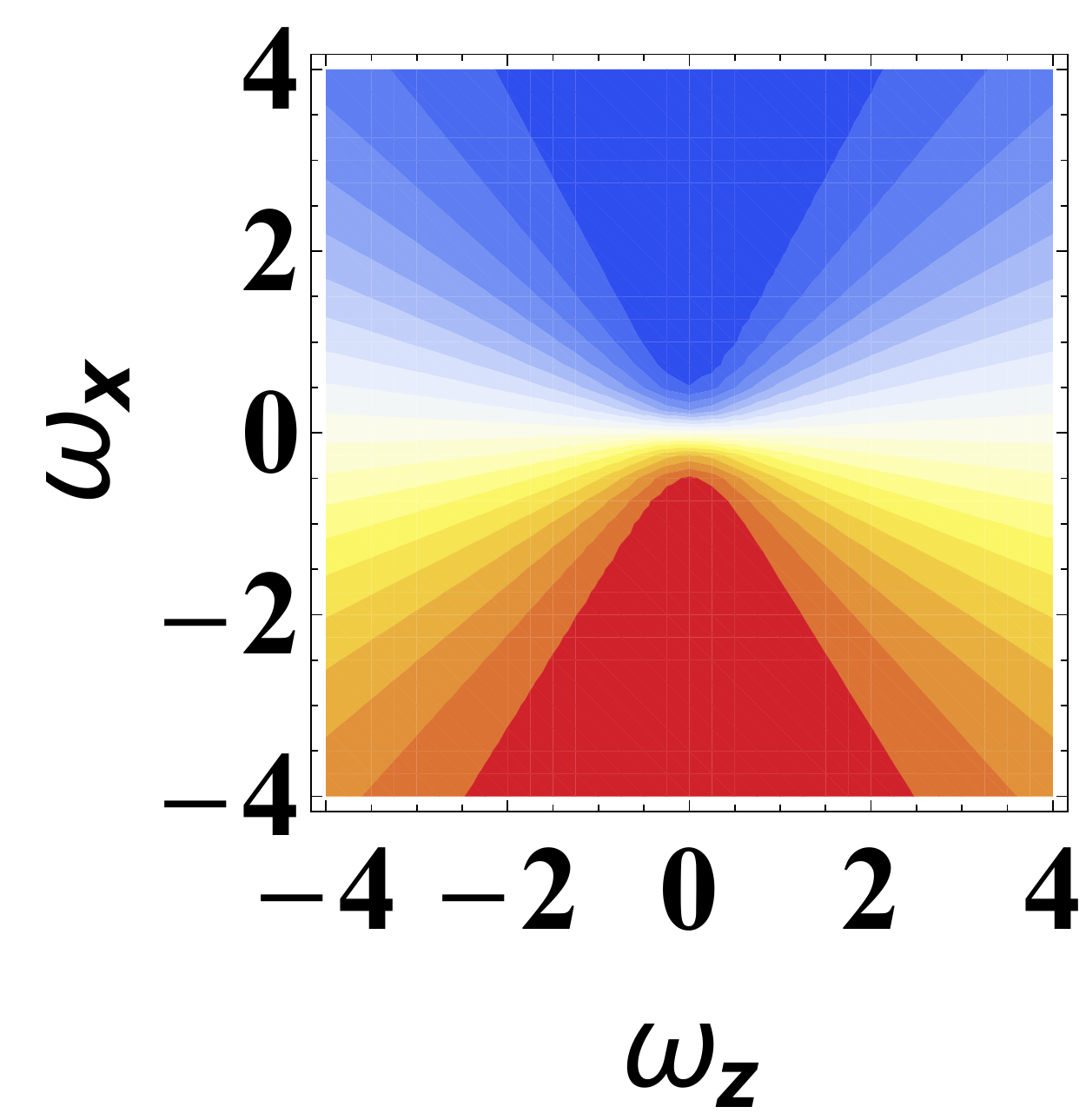}
\label{fig:1g}}
\subfigure[$\theta=+3\pi/8$]{
\includegraphics[width=3cm]{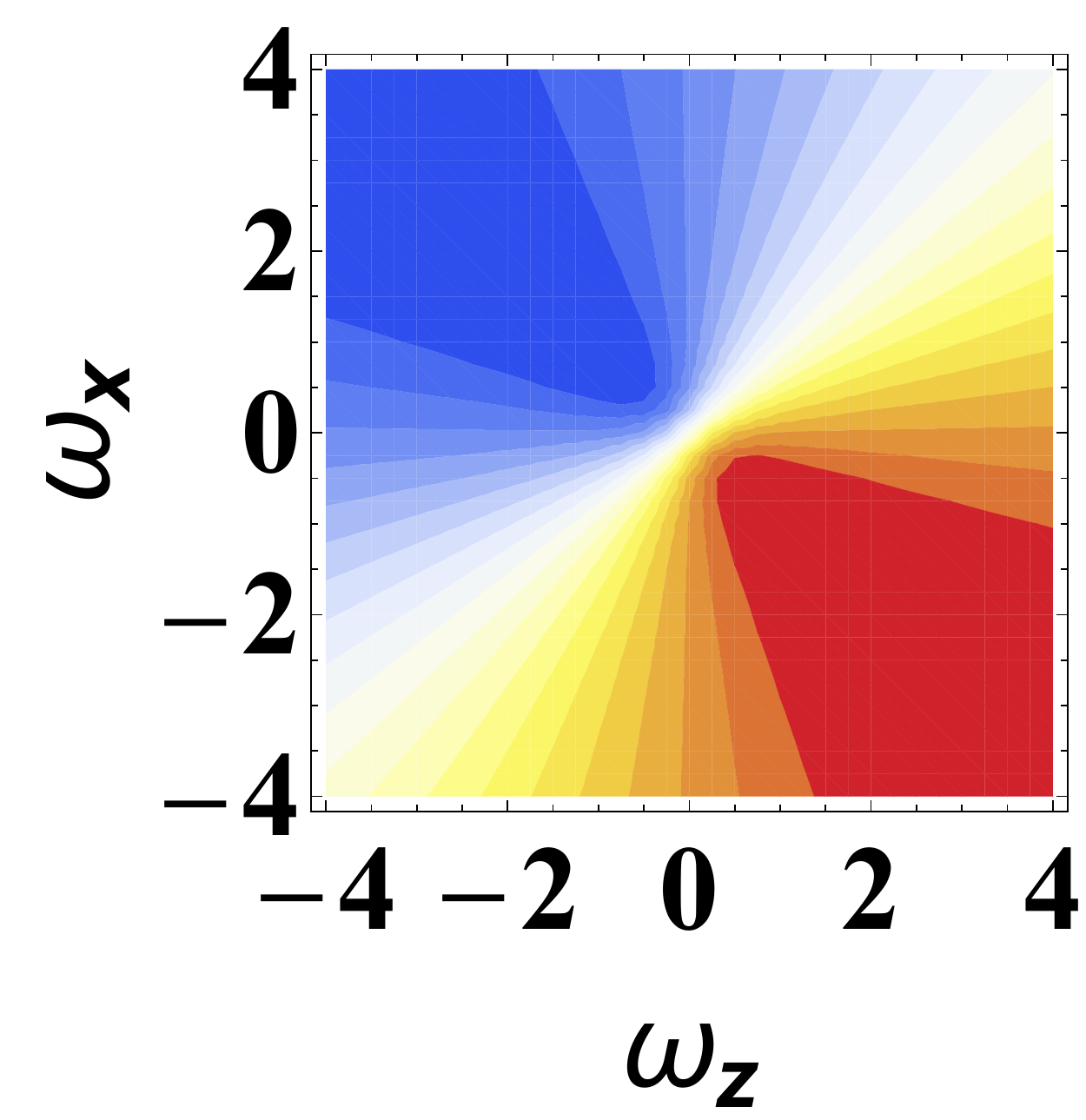}
\label{fig:1h}}
\caption{The cross section of fidelity defined in eq.(\ref{eq:fidelity}), with $\omega_1=-\omega_2=\omega_z$, $\omega_3=-\omega_4=\omega_x$, $\omega_5=-\omega_6=0$ and $\epsilon=0.1$.\label{fig:1}}
\end{figure}

\begin{figure}[h]
\centering
\includegraphics[width=.7cm]{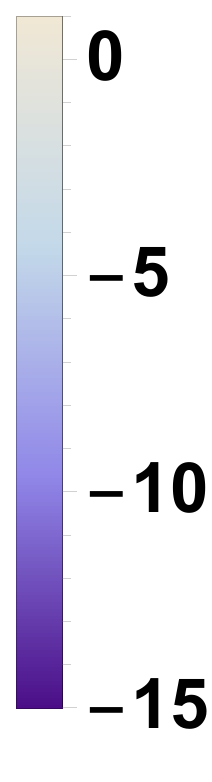}
\includegraphics[width=3cm]{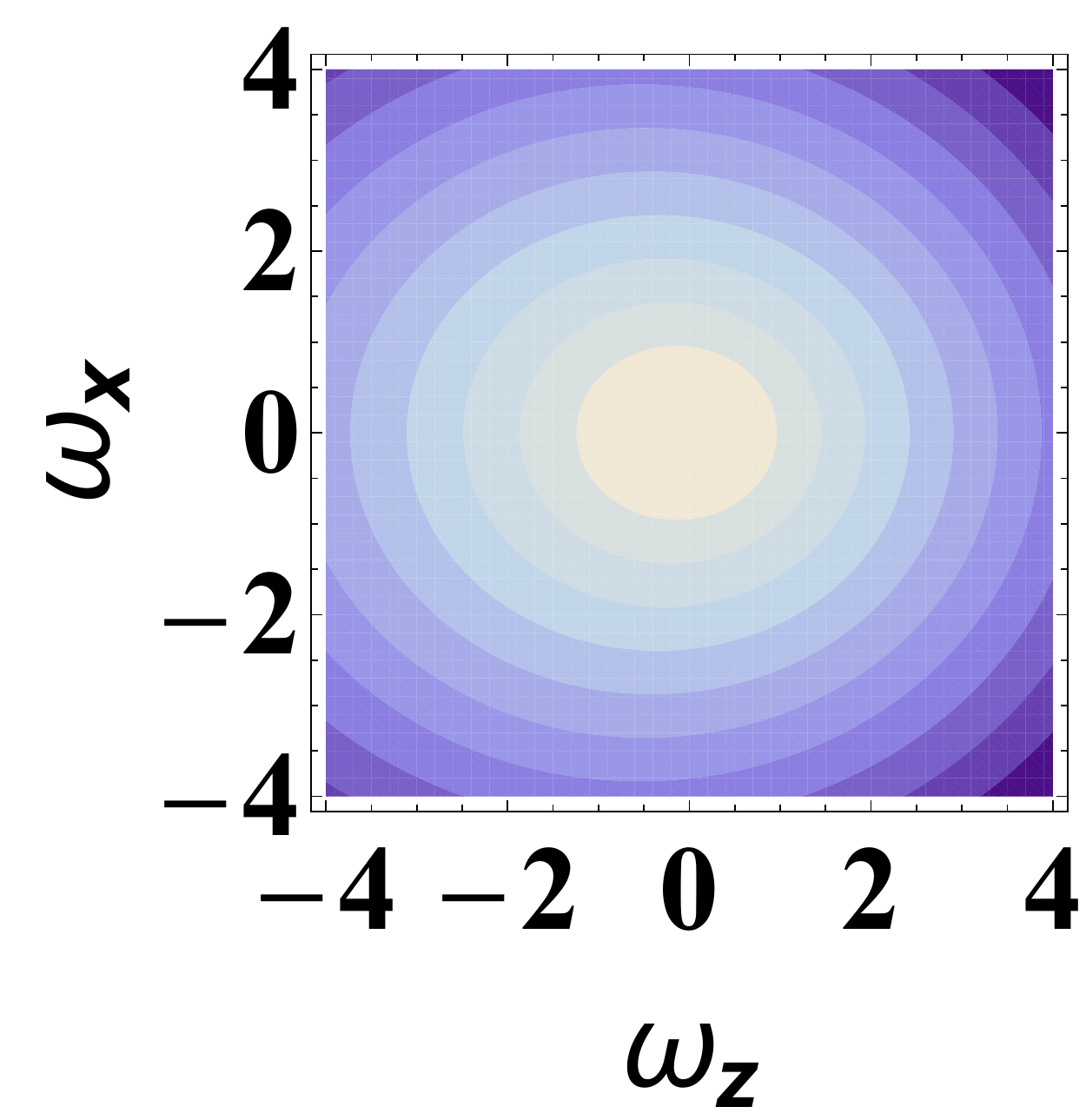}
\caption{$(\omega_z,\omega_x)$-dependence of $\omega_0$, with $\omega_1=-\omega_2=\omega_z$, $\omega_3=-\omega_4=\omega_x$, $\omega_5=-\omega_6=0$ and $\epsilon=0.1$.\label{fig:2}}
\end{figure}

Let us show how the parametrization works. As the simplest example, we look closely at the case where the target state is given in the two dimensional Hilbert space. For the state, we consider a set in eq.(\ref{eq:models}) with
\begin{equation}
{\mathcal M}=
\left\{
\begin{array}{ll}
\hat{\rho}_1=
\displaystyle{\frac{1}{2}}
\left(
\begin{array}{cc}
2-\epsilon& 0\\
0& \epsilon
\end{array}
\right),&
\hat{\rho}_2=
\displaystyle{\frac{1}{2}}
\left(
\begin{array}{cc}
\epsilon & 0\\
0& 2-\epsilon
\end{array}
\right)\\
\hat{\rho}_3=
\displaystyle{\frac{1}{2}}
\left(
\begin{array}{cc}
1 & 1-\epsilon\\
1-\epsilon & 1
\end{array}
\right),&
\hat{\rho}_4=
\displaystyle{\frac{1}{2}}
\left(
\begin{array}{cc}
1 & -(1-\epsilon)\\
-(1-\epsilon) & 1
\end{array}
\right)\\
\hat{\rho}_5=
\displaystyle{\frac{1}{2}}
\left(
\begin{array}{cc}
1 & -i(1-\epsilon)\\
i(1-\epsilon) & 1
\end{array}
\right),&
\hat{\rho}_6=
\displaystyle{\frac{1}{2}}
\left(
\begin{array}{cc}
1 & i(1-\epsilon)\\
-i(1-\epsilon) & 1
\end{array}
\right)
\end{array}
\right\}
\label{eq:6 models}
\end{equation}
as models where $\epsilon$ is a small positive number introduced to make $\hat{\eta}_i$ in eq.(\ref{eq:etas}) mathematically well defined.
For example, the models given by $\hat{\rho}_1$ and $\hat{\rho}_2$ represent the (almost) eigenstates of
\begin{equation}
\hat{\sigma}_z:=
\left(
\begin{array}{cc}
1 & 0\\
0 & -1
\end{array}
\right)
\end{equation}
with eigenvalue $+1$ and $-1$ respectively. Other models $(\hat{\rho}_3, \hat{\rho}_4)$ and $(\hat{\rho}_5, \hat{\rho}_6)$ have similarly represent the (almost) eigenstates of 
\begin{equation}
\hat{\sigma}_x:=
\left(
\begin{array}{cc}
0 & 1\\
1 & 0
\end{array}
\right),\quad
\hat{\sigma}_y:=
\left(
\begin{array}{cc}
0 & -i\\
i & 0
\end{array}
\right)
\end{equation}
respectively. With these models, we have seven parameters $(\omega_0,\omega_1,\omega_2,\omega_3,\omega_4,\omega_5,\omega_6)$ to represent $\hat{\rho}({\bm \omega})$ in eq.(\ref{eq:parameterization}). Note that $\omega_0$ is determined by the normalization condition for $\hat{\rho}({\bm \omega})$, {\it i.e.},
\begin{equation}\label{eq:normalization condition w0}
\omega_0=-\ln {\rm tr}\left(\exp\left(\sum_{i=1}^6 \omega_i \hat{\eta}_i\right)\right).
\end{equation}
To numerically check how the parameterization work, let us consider an example with a pure state
\begin{equation}
\hat{\mu}_\theta:=\left(
\begin{array}{cc}
\cos^2 \theta & \sin\theta\cos\theta\\
\sin\theta\cos\theta & \sin^2\theta
\end{array}
\right)
\end{equation}
as the target state. Computing the fidelity\cite{doi:10.1080/09500349414552171}
\begin{equation}\label{eq:fidelity}
F(\hat{\mu}_\theta,\hat{\rho}({\bm \omega})):=\left[{\rm tr}\left(\sqrt{\sqrt{\hat{\mu}_\theta}\hat{\rho}({\bm\omega})\sqrt{\hat{\mu}_\theta}}\right)\right]^2,
\end{equation}
as shown in Figure \ref{fig:1}, we can be sure that there exists a parameter region where the fidelity achieves almost $1$ for each $\theta$. $(\omega_z,\omega_x)$-dependence of $\omega_0$ estimated by eq.(\ref{eq:normalization condition w0}) is shown in Figure \ref{fig:2}.
 

For a general target state $\hat{\mu}$ and the parameterized state $\hat{\rho}({\bm \omega})$, the negatively signed fidelity $$-F(\hat{\mu},\hat{\rho}({\bm \omega}))$$ can be an objective function to find parameters ${\bm \omega}$ minimizing the function. Our aim in this article, however, is to introduce a physically implementable objective function as Hamiltonian to solve the finding parameters problem by applying the quantum annealing computation. To do so, instead of the fidelity since we do not know how to physically implement it although it might be possible, we employ the relative entropy (more precisely, the quantity in eq.(\ref{eq:objective function D})) as an objective function, {\it i.e.}, the metric to find the optimal parameters. In the following sections, we discuss some implications of the metric and how to physically implement it.

\section{Relative Entropy as Metric}
Here, we give the implications of our use of the relative entropy as the metric. The relative entropy is a quantum extension of the notion of Kullback-Leibler distance\cite{kullback1951}. In general, the relative entropy of state $\hat{\rho}$ with respect to $\hat{\mu}$ defined by
\begin{equation}
R(\hat{\mu};\hat{\rho}):={\rm tr}\left(\hat{\mu}\log \hat{\mu}\right)-{\rm tr}\left(\hat{\mu}\log\hat{\rho}\right)
\end{equation}
is proven to be bounded below by $0$, and is $0$ if and only if $\hat{\rho}=\hat{\mu}$. Applying the above to $\hat{\rho}({\bm \omega})$ and $\hat{\mu}$, and focussing on its ${\bm \omega}$-dependence only, we employ $D(\hat{\mu};\hat{\rho}({\bm \omega}))$ in eq.(\ref{eq:objective function D}) as the metric. The quantity is bounded by the Shannon entropy of the target state, and in the case where both models in ${\mathcal M}$ and the parameter range of ${\bm \omega}$ are appropriately given, the bound will be appropriately achieved.

As shortly mentioned in the introduction section, however, our aim in this article is to introduce a Hamiltonian corresponding to $D(\hat{\mu};\hat{\rho}({\bm \omega}))$ with the degree of freedoms of $\hat{\bm \omega}$. For the purpose, there are some points we need to take care. First point is that, when ${\bm \omega}$ is treated as free parameters, the expression in eq.(\ref{eq:parameterization}) does not automatically imply that $\hat{\rho}(\bm{\omega})$ is state. That is because 
$${\rm tr}\left(\hat{\rho}({\bm \omega})\right)$$
can be any positive number. Taking account into this point, we define a metric by adding a constraint term as
\begin{equation}\label{eq:parametrization2}
E({\bm \omega}):=D(\hat{\mu};\hat{\rho}({\bm \omega}))+\alpha \left({\rm tr}\left(\hat{\rho}({\bm \omega})\right)-1\right)^2
\end{equation}
with $\alpha >0$. The second point is the parameter range of ${\bm \omega}$. Since Ising model will be supposed as a system on which  the annealing computation will be implemented, the parameter ${\bm \omega}$ needs to be discretized parameters. Due to the discretization, even ${\bm \omega}$ minimizing $E({\bm \omega})$ is not necessarily giving a state when substituted into $\hat{\rho}({\bm \omega})$. For this reason, we need to introduce a normalization;
\begin{equation}\label{eq:normalized state}
\hat{\rho}_R({\bm \omega}):=\frac{\hat{\rho}({\bm \omega})}{{\rm tr}\left(\hat{\rho}\left({\bm \omega}\right)\right)}
\end{equation}
to represent the result of the tomography as a state. (The normalization can be obtained after solving ${\bm \omega}$ minimizing $E({\bm \omega}).$)

\begin{figure}[h]
\centering
\subfigure[Metric vs Entropy of target state $\hat{\mu}$]{
\includegraphics[width=7cm]{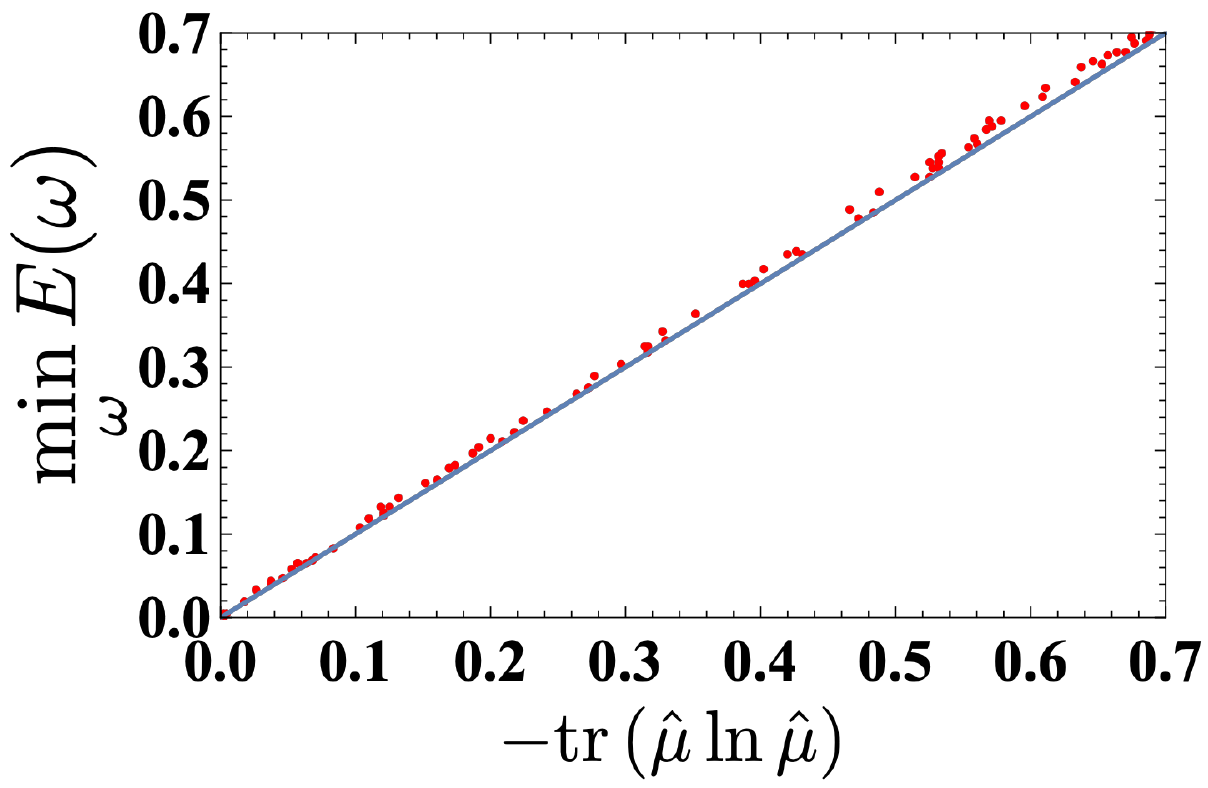}
\label{fig:3a}}~~
\subfigure[Fidelity vs Entropy of target state $\hat{\mu}$]{
\includegraphics[width=7cm]{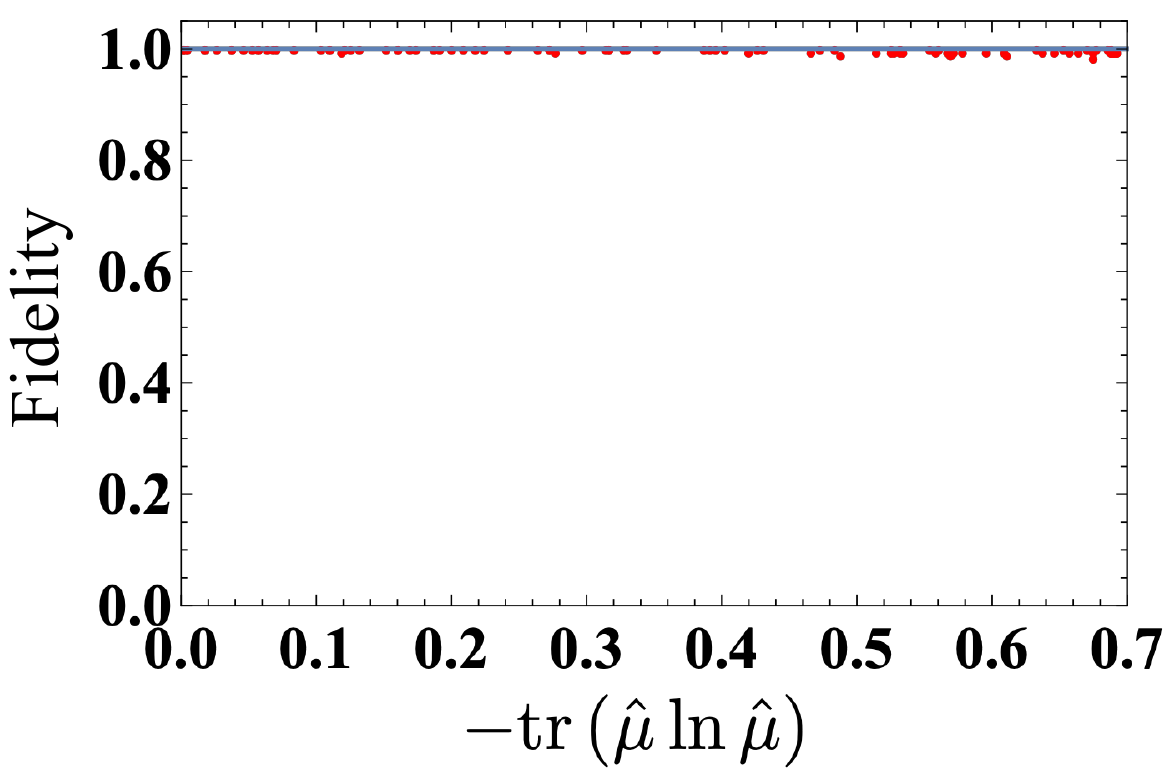}
\label{fig:3b}}
\caption{On 100 target states, minimizing argument problem for $E({\bm \omega})$ is numerically examined with $\alpha=100$ by the random search method over variables $(\omega_1,\omega_2,\omega_3,\omega_4,\omega_5,\omega_6)$ associated with ${\mathcal M}$ in eq.(\ref{eq:6 models}) with $\epsilon=0.1$. (Each $\omega_i$ is discretized as a $12$-bit signed real number. $\omega_0$ is fixed to $0$ as a tentative choice.) Each red point corresponds to one target state randomly generated.\label{fig:3}}
\end{figure}

Let us numerically check efficiency of eqs.(\ref{eq:parametrization2}) and (\ref{eq:normalized state}). Preparing $\hat{\mu}$ in the two dimensional Hilbert space randomly, we compute the minimizing argument of eq.(\ref{eq:parametrization2}) by the random search method. Substituting ${\bm \omega}$ for the argument, we estimate the metric in eq.(\ref{eq:parametrization2}) and the fidelity between $\hat{\mu}$ and the normalized state in eq.(\ref{eq:normalized state}), as plotted in Figure\ref{fig:3}. We find that the metric actually works well for the purpose of the quantum tomography.

On these premises, introducing the degree of freedoms $\hat{\bm\omega}$, we will show that we can construct an effective Hamiltonian such as
\begin{equation}\label{eq:effective Hamiltonian}
\hat{H}_{eff}(\hat{\bm \omega})|{\bm \omega}\rangle=E({\bm\omega}) |{\bm \omega}\rangle
\end{equation}
in the next section. 

\section{Construction of Effective Hamiltonian}
Here, we theoretically investigate a way to physically implement an effective Hamiltonian described in eq.(\ref{eq:effective Hamiltonian}).
We assume that copies of target state $\hat{\mu}$ on Hilbert space ${\mathcal H}$ are physically available as much as required. For the sake of convenience, we write down as
\begin{equation}\label{eq:copy}
\hat{\mu}^{(s)}\in {\mathcal H}^{(s)}
\end{equation}
for the $s$-th copy of $\hat{\mu}$ so as each ${\mathcal H}^{(s)}$ is isomorphic to ${\mathcal H}$. (Although we will see how we use the copy later on, we continue our discussion supposing a fixed $s$ for the moment.)
Remember that models $\{\hat{\rho}_i\}_{i=1}^m$ and their associated operators $\{\hat{\eta}_i\}_{i=1}^m$ are defined in the same Hilbert space as ${\mathcal H}^{(s)}$. To make this point clear, we add index $s$ to them as
\begin{equation}
\{\hat{\rho}_i^{(s)}\}_{i=1}^m,
\quad\mbox{and}\quad
\{\hat{\eta}_i^{(s)}\}_{i=1}^m.
\end{equation}
Similarly, we introduce the identity operator $\hat{I}_{{\mathcal H}^{(s)}}$ as the identity operator on ${\mathcal H}^{(s)}$.

Besides these operators (and Hilbert spaces), to introduce the degree of freedom corresponding to $\hat{\bm \omega}=(\hat{\omega}_0,\cdots,\hat{\omega}_m)$, we additionally introduce Hilbert space
\begin{equation}
{\mathcal H}_{{\bm \Omega}}:=\bigotimes_{i=0}^{m}{\mathcal H}_{\Omega_m}
\end{equation}
where each $\hat{\omega}_i$ lives in ${\mathcal H}_{\Omega_i}$. Reflecting the discretization referred in the previous section, we suppose that each ${\mathcal H}_{\Omega_i}$ is isomorphic to ${\mathbb C}^{\otimes n}$. With this setting, we can safely introduce $\hat{\omega}_i$ in Hilbert space ${\mathcal H}_{\Omega_i}$, corresponding to the binary representation of $\omega_i$ in $n/2$ digits. In the following, for such $\hat{\bm \omega}$, we show that $\hat{H}_{eff}(\hat{\bm\omega})$ in eq.(\ref{eq:effective Hamiltonian}) with $E({\bm \omega})$ given in eq.(\ref{eq:parametrization2}) can be physically implemented. 

\subsection*{Corresponding to the first term of $E({\bm\omega})$ in eq.(\ref{eq:parametrization2})}
Now, let us consider an operator
\begin{equation}\label{eq:grnd Hamiltonian1}
\hat{H}_D^{(s)}:=\sum_{i=0}^m\hat{\omega}_i\otimes \hat{\eta}_i^{(s)}
\end{equation}
defined on ${\mathcal H}_{{\bm \Omega}}\otimes{\mathcal H}^{(s)}$. Suppose if the state on the compound Hilbert space is separable state
\begin{equation}\label{eq:separable state}
\hat{\chi}_t\otimes\hat{\mu}^{(s)}.
\end{equation}
Then, the state is evolved by the operator in eq.(\ref{eq:grnd Hamiltonian1}) is govern by von Neumann equation 
\begin{equation}\label{eq:total dynamics}
\frac{d}{dt} \hat{\chi}_t\otimes\hat{\mu}^{(s)}=
-i
\left[
\hat{H}_D^{(s)},\hat{\chi}_t\otimes\hat{\mu}^{(s)}
\right]
\end{equation}
and
\begin{equation}\label{eq:1st term}
\frac{d}{dt}\hat{\chi}_t=
-i
\left[ {\rm tr}_{{\mathcal H}^{(s)}}\left(
\hat{H}_D^{(s)}\hat{\mu}^{(s)}\right),\hat{\chi}_t
\right]
\end{equation}
where ${\rm tr}_{{\mathcal H}^{(s)}}$ denotes the partial trace over Hilbert space ${\mathcal H}^{(s)}$. 
Notice that
\begin{eqnarray}
{\rm tr}_{{\mathcal H}^{(s)}}\left(
\hat{H}_D^{(s)}\hat{\mu}^{(s)}\right)&=&-{\rm tr}_{\mathcal H}\Big(\hat{\mu}\ln \hat{\rho}({\bm \omega})\Big)|_{{\bm \omega}\rightarrow\hat{\bm \omega}\in {\mathcal H}_{{\bm \Omega}}}\nonumber\\
&=&
D(\hat{\mu};\hat{\rho}({\bm \omega}))|_{{\bm \omega}\rightarrow\hat{\bm \omega}\in {\mathcal H}_{{\bm \Omega}}}
\end{eqnarray}
which corresponds to the first term of $E({\bm\omega})$ in eq.(\ref{eq:parametrization2}). 

\bigskip

\noindent
{\bf Note 1:}
Eq.(\ref{eq:1st term}) can be justified only in the case where the assumption in eq.(\ref{eq:separable state}) holds. The evolution by eq.(\ref{eq:total dynamics}), however, {\it does not} generally maintain the form of eq.(\ref{eq:separable state}). Expanding both solutions of eq.(\ref{eq:total dynamics}) and eq.(\ref{eq:1st term}) by small time interval $\delta t$, one can find that the dynamics by eq.(\ref{eq:total dynamics}) follows the dynamics by eq.(\ref{eq:1st term}) up to the first order of $\delta t$. Thus, we need to supply a new copy of $\hat{\mu}$ as the interacting partner after every time evolutions by $\delta t$ so as we can actually obtain the dynamics described in eq.(\ref{eq:total dynamics}) for a finite time with the error of $O(\delta t)$. This is how we use the copies described in eq.(\ref{eq:copy}). Similar argument can be found in \cite{cmki2019}. We will come to the point of the choice of $\delta t$ in the last part of this article.

\subsection*{Corresponding to the second term of $E({\bm\omega})$ in eq.(\ref{eq:parametrization2})}
Concerning the second term, we can physically implement the term by the following trick: As is described in \cite{Lloyd2014}, one can generally let any quantum state work as a Hamiltonian. In concrete, when state $\tilde{\Xi}$ can be freely prepared in a Hilbert space which is isomorphic to ${\mathcal G}$, we can physically implement the time evolution of state $\hat{\Xi}_t$ in ${\mathcal G}$ that follows
\begin{equation}\label{eq:25}
\frac{d}{dt}\hat{\Xi}_t=-i\lambda \left[\tilde{\Xi},\hat{\Xi}_t\right].
\end{equation}
with a constant $\lambda$. Now, let us consider the case where
\begin{enumerate}
\item $\tilde{\Xi}$ is the thermal state in terms of Hamiltonian $-\hat{H}_D^{(u')}$ in temperature $\beta^{-1}$, {\it i.e.},
\begin{equation}\label{eq:Z1}
\tilde{\Xi}=\frac{1}{Z_1(\beta)}\exp\left(\beta \hat{H}_D^{(u')}\right),~\mbox{with}~
Z_1(\beta)={\rm tr}\left(\exp\left(\beta \hat{H}_D^{(u')}\right)\right)
\end{equation}
where $\hat{H}_D^{(u')}$ defined in eq.(\ref{eq:grnd Hamiltonian1}) is an operator on Hilbert space ${\mathcal H}_{\bm \Omega}\otimes{\mathcal H}^{(u')}$.

\item $\hat{\Xi}_t$ is prepared in a separable state:
\begin{equation}\label{eq:initial state2}
\hat{\Xi}_t:=\hat{\chi}_t\otimes\frac{\hat{I}_{{\mathcal H}^{(u')}}}{\dim {\mathcal H}^{(u')}}.
\end{equation}
\end{enumerate}
Putting $\lambda=-2 \alpha Z_1(\beta)$ and $\beta=1$ (with an appropriate unit), the dynamics in (\ref{eq:25}) can be reduced as 
\begin{equation}\label{eq:2c term}
\frac{d}{dt}\hat{\chi}_t=+2\alpha i\left[\hat{C}_1(\hat{\bm \omega}),\hat{\chi}_t\right]
\end{equation}
where
\begin{equation}
\hat{C}_1(\hat{\bm \omega}):={\rm tr}_{{\mathcal H}^{(u')}}\left(\exp\left(\hat{H}_D^{(u')}\right)\right).
\end{equation}
Notice that $Z_1(\beta)$ in eq.(\ref{eq:Z1}) is independent from target state $\hat{\mu}$ and that it is possible to compute it in advance of the tomographic process. Similarly, by considering the case where
\begin{enumerate}
\item $\tilde{\Xi}$ is the thermal state in terms of Hamiltonian $-(\hat{H}_D^{(u'')}+\hat{H}_D^{(u''')})$ in temperature $\beta^{-1}$, {\it i.e.},
\begin{equation}\label{eq:Z1}
\tilde{\Xi}=\frac{1}{Z_2(\beta)}\exp\left(\beta \left(\hat{H}_D^{(u'')}+\hat{H}_D^{(u''')}\right)\right),~\mbox{with}~
Z_2(\beta)={\rm tr}\left(\exp\left(\beta \left(\hat{H}_D^{(u'')}+\hat{H}_D^{(u''')}\right)\right)\right)
\end{equation}
\item $\hat{\Xi}_t$ is prepared in a separable state:
\begin{equation}\label{eq:initial state3}
\hat{\Xi}_t:=\hat{\chi}_t\otimes\frac{\hat{I}_{{\mathcal H}^{(u'')}}}{\dim {\mathcal H}^{(u'')}}\otimes\frac{\hat{I}_{{\mathcal H}^{(u''')}}}{\dim {\mathcal H}^{(u''')}}.
\end{equation}
\end{enumerate}
Putting $\lambda=\alpha Z_2(\beta)$ and $\beta=1$ (with an appropriate unit), the dynamics in (\ref{eq:25}) can be reduced as 
\begin{equation}\label{eq:c^2 term}
\frac{d}{dt}\hat{\chi}_t=-\alpha i\left[\hat{C}_2(\hat{\bm \omega}),\hat{\chi}_t\right]
\end{equation}
where
\begin{eqnarray}
\hat{C}_2(\hat{\bm \omega})&:=&{\rm tr}_{{\mathcal H}^{(u'')}}\left(\exp\left(\hat{H}_D^{(u'')}\right)\right)~
{\rm tr}_{{\mathcal H}^{(u''')}}\left(\exp\left(\hat{H}_D^{(u''')}\right)\right)\nonumber\\
&=&\hat{C}_1(\hat{\bm \omega})^2.
\end{eqnarray}
Notice that $Z_2(\beta)$ in eq.(\ref{eq:Z1}) is also possible to compute in advance of the tomographic process.
Combining eqs.(\ref{eq:2c term}), (\ref{eq:c^2 term}) and a time evolution by trivial Hamiltonian $\alpha\hat{I}_{{\mathcal H}_{\bm \Omega}}$, we can implement the time evolution following
\begin{equation}\label{eq:2nd term}
\frac{d}{dt}\hat{\chi}_t=-i\left[\alpha \left(\hat{C}_1(\hat{\bm \omega})-\hat{I}_{{\mathcal H}_{\bm \Omega}}\right)^2,\hat{\chi}_t\right]
\end{equation}
that corresponds to the second term in eq.(\ref{eq:parametrization2}).
\bigskip

\noindent
{\bf Note 2:} Similarly to the argument in the Note 1 above, to justify eqs.(\ref{eq:2c term}) , (\ref{eq:c^2 term}) and (\ref{eq:2nd term}) up to a finite time evolution, states
\begin{equation}
\frac{\hat{I}_{{\mathcal H}^{(u')}}}{\dim {\mathcal H}^{(u')}},~
\frac{\hat{I}_{{\mathcal H}^{(u'')}}}{\dim {\mathcal H}^{(u'')}},~\mbox{and}~
\frac{\hat{I}_{{\mathcal H}^{(u''')}}}{\dim {\mathcal H}^{(u''')}}
\end{equation}
in eqs.(\ref{eq:initial state2}) and (\ref{eq:initial state3}) are supposed to be newly supplied after every short time evolutions by $\delta t$. Unlike the case of Note 1, these states can be prepared independently from target state $\hat{\mu}$.

Combining eqs.(\ref{eq:1st term}) and (\ref{eq:2nd term}) under the circumstances referred in Note 1 and Note 2, we achieve the dynamics
\begin{equation}\label{eq:original one}
\frac{d}{dt}\hat{\chi}_t=-i\left[\hat{H}_{eff}(\hat{\bm \omega}),\hat{\chi}_t\right]
\end{equation}
where $\hat{H}_{eff}(\hat{\bm \omega})$ is given in eq.(\ref{eq:effective Hamiltonian}).

\subsection*{Application to Quantum Annealing Computation}
\begin{figure}[h]
\centering
\includegraphics[width=14cm]{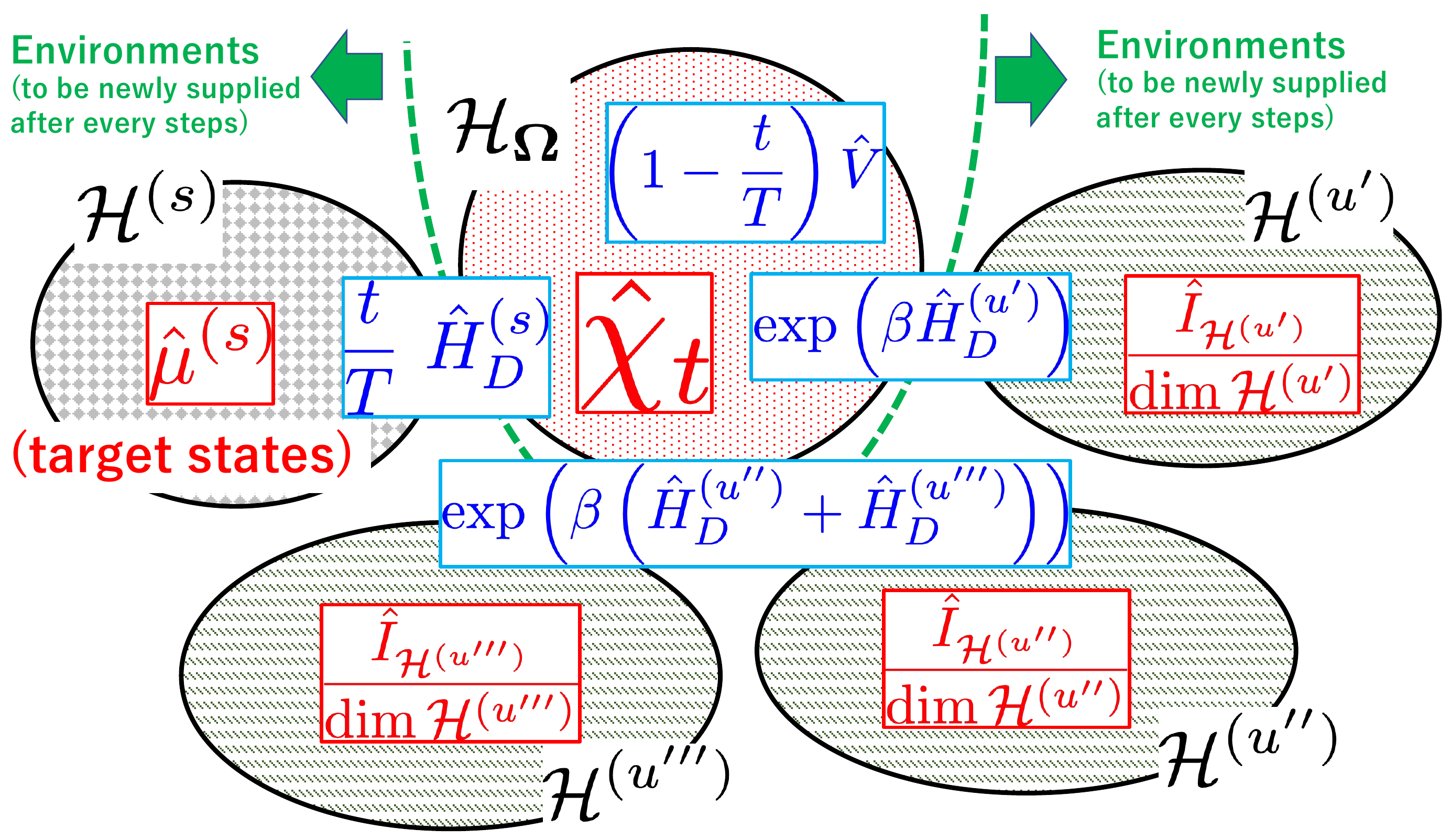}
\caption{Our construction of $\hat{H}_{QA}(t)$ as an effective Hamiltonian on ${\mathcal H}_{\bm \Omega}$. The blue items represent interaction Hamiltonians among systems on each Hilbert space. (Concerning the way to implement the "exp" type interaction Hamiltonians, see eq.(\ref{eq:25}) and \cite{Lloyd2014} for some details.) The red items represent states in each Hilbert space. The states in ${\mathcal H}^{(s)}$, ${\mathcal H}^{(u')}$, ${\mathcal H}^{(u'')}$, and ${\mathcal H}^{(u''')}$ need to be newly supplied one after another after every short time evolutions by $\delta t$.\label{fig:4}}
\end{figure}

Let us consider an application of the above idea to the quantum annealing computation. That can be done by additionally introducing a driving Hamiltonian $$\left(1-\frac{t}{T}\right)\hat{V}$$ on ${\mathcal H}_{\bm \Omega}$, and by replacing $\hat{H}_D^{(s)}$ in eq.(\ref{eq:copy}) by a time dependent one as $$\frac{t}{T} \hat{H}_D^{(s)}.$$ Then, instead of eq.(\ref{eq:original one}), we obtain
\begin{equation}\label{eq:QA one}
\frac{d}{dt}\hat{\chi}_t=-i\left[\hat{H}_{QA}(t),\hat{\chi}_t\right]
\end{equation}
where
\begin{equation}
\hat{H}_{QA}(t):=\frac{t}{T} \hat{H}_{eff}(\hat{\bm \omega})+\left(1-\frac{t}{T}\right)\hat{V}.
\end{equation}
Notice that the above holds only under the circumstances referred in Note 1 and Note 2. The construction of $\hat{H}_{QA}(t)$ on ${\mathcal H}_{\bm \Omega}$ is summarized in Figure \ref{fig:4}.

By choosing $\hat{\chi}_{t=0}$ to be the ground state of $\hat{V}$, $\hat{\chi}_t$ converges on the ground state of $\hat{H}_{eff}(\hat{\bm \omega})$ as is desired when $T$ is appropriately chosen as is generally required in the quantum annealing process. Since the error of $O(\delta t^2)$ can be accumulated for every short time evolutions by $\delta t$,
$$O\left(\delta t^2 \frac{T}{\delta t}\right)=O(\delta t T)$$ must be constant so as the error can be constantly bounded. (Remember that $\delta t$ and $T$ in this article are dimensionless quantities so as our effective Hamiltonian is.) Thus, $\delta t$ is determined to be $O(T^{-1})$. The required number of the copies of the target state $\hat{\mu}$ can be also determined as $O(T^2)$. Notice that the appropriate $T$ itself generally depends on $\dim {\mathcal H}_{\bm \Omega}$, choice of $\hat{V}$, and so on. Estimation of the appropriate $T$ itself with some concrete situations needs to be addressed in future works.


\end{document}